%% file: Template.tex
\documentclass{article}
\usepackage{spconf,amsmath,graphicx,hyperref}
\usepackage{amssymb}
\usepackage{booktabs}
\usepackage{xcolor}

\title{Melody Guidance for Audio Diffusion Models Without Retraining}
\twoauthors
 {A. Author-one, B. Author-two\sthanks{Thanks to XYZ agency for funding.}}
	{School A-B\\
	Department A-B\\
	Address A-B}
 {C. Author-three, D. Author-four\sthanks{The fourth author performed the work
	while at ...}}
	{School C-D\\
	Department C-D\\
	Address C-D}
    \name{Cristóbal Andrade$^{\star }$ \qquad Sebastian J. Schlecht$^{\star}$ \qquad Ben Maman$^{\dagger}$ \qquad Meinard Müller$^{\dagger}$}
  
  \address{$^{\star}$Multimedia Comms. \& Signal Process., Friedrich-Alexander-Universität Erlangen-Nürnberg, Germany\\
      $^{\dagger}$International Audio Laboratories Erlangen, Erlangen, Germany}
\begin{document}
\ninept
\maketitle
%
% \begin{abstract}
% Melody control is central to many audio generation tasks, existing approaches achieve it by training generative models directly on symbolic conditioning signals such as F0 or MIDI. We present a training-free method for guiding \emph{unconditionally} trained diffusion models toward a target melody at inference time, without any additional training or melody-specific model conditioning. We formulate melody conditioning as an inverse problem, using a differentiable pitch estimator as the degradation operator within a diffusion posterior sampling framework, and introduce a melody-informed warm initialization that stabilizes generation against the underdetermined nature of pitch-only guidance. Our method is model-agnostic, applying without modification to both waveform- and latent-domain diffusion models. Across four models spanning piano, guitar, and singing voice, our approach achieves melody adherence competitive with dedicated conditional baselines, while requiring no melody-specific training.
% \end{abstract}

\begin{abstract}
    Controlling melody in audio generation normally requires training a conditional generative model on audio paired with pitch curves or MIDI notes. We instead guide an existing unconditional diffusion model, originally trained without melody inputs, to follow a target melody without retraining or modifying it. During inference, a differentiable pitch detector steers the sampled audio toward the desired pitches, while the diffusion model keeps it close to the distribution of realistic audio learned during training. Our main contribution is a sampling method that balances these two directions, providing melody control while preserving audio quality. To improve stability, we initialize sampling with a simple synthesized version of the target melody rather than pure noise. Experiments on piano, guitar, and singing voice show melody accuracy comparable to conditional models trained specifically for melody control. Our method therefore adds melody control to existing audio diffusion models without requiring paired melody data or task-specific retraining.

\end{abstract}

\section{Introduction}
Melody, the sequence of pitches unfolding over time, is one of the fundamental organizing elements of music, and precise control over it underlies a broad range of audio generation tasks. Fundamental frequency (F0) and melody conditioning have been used to drive singing voice synthesis and conversion~\cite{liu2021diffsvc, MamanEtAl26_DiffusionVoiceConversion_DAFx}, vocoder-based speech synthesis~\cite{MoriseYO16}, controllable sound generation from time-varying sonic imitations~\cite{garcía2025sketch2soundcontrollableaudiogeneration}, single- and multi-instrument music synthesis~\cite{jonason2024ddspguitar, hawthorne2022multi}, and timbre transfer~\cite{comanducci2023timbretransferusingimagetoimage}. 

Existing approaches implement melody control by conditioning a generative model on a symbolic or acoustic melody representation during training. Common conditioning signals include F0 curves injected through dedicated pitch encoders or concatenated with content embeddings~\cite{liu2021diffsvc, MoriseYO16}, MIDI processed into note- or performance-level embeddings via hierarchical or sequence encoders~\cite{MamanEtAl26_DiffusionVoiceConversion_DAFx, jonason2024ddspguitar, hawthorne2022multi}, and spectrograms used directly as conditioning signals~\cite{comanducci2023timbretransferusingimagetoimage}. In all cases, melody control is learned and requires a model trained or fine-tuned with melody-specific conditioning.

In this work, we seek to control the melody of an unconditional diffusion model at inference time, without melody-specific training or modifications to the pretrained model. We formulate this as an inverse problem, where the goal is to recover a signal from a degraded measurement of it. In our setting, we treat the F0 representation $\boldsymbol{y}$ as a degraded measurement of the audio signal, obtained through a differentiable pitch estimator. Given a desired melody, we then seek plausible audio whose estimated pitch contour matches $\boldsymbol{y}$.

Diffusion posterior sampling (DPS) \cite{chung2023diffusion} provides a natural mechanism for this objective. It combines the diffusion model’s prior, which promotes realistic audio, with a likelihood term that guides generation toward the target melody. Closest to our work, \cite{ye2026pitchclass} steers an unconditional diffusion model's pitch via a trained latent pitch estimator. In contrast, we require no auxiliary training. To our knowledge, no prior work has formulated melody-guided generation for unconditional diffusion models in this way. This setting is challenging because, unlike the well-conditioned and approximately linear operators used in most audio inverse problems \cite{moliner2023solving}, F0 estimation is nonlinear and severely underdetermined. It maps a high-dimensional waveform to a compressed single pitch representation, so many acoustically distinct signals can share the same contour. Additionally, the Gaussian likelihood approximation that DPS relies on is not valid. To support DPS, warm initialization \cite{andrade2026audio}, a technique that starts reverse diffusion from a structured guide signal rather than from pure noise, can be used. This allows to begin inference closer to a melodically plausible signal, reducing the corrective burden placed on the guidance term and by that stabilizing sampling.
% \ben{
% To address this setting, we present a model-agnostic, training-free method for guiding unconditional diffusion models toward a target F0 trajectory. Our method uses a differentiable pitch estimator as the measurement operator within DPS. To mitigate the distributional drift commonly observed with DPS~\cite{chung2022improving}, we employ a melody-informed warm initialization based on an approximate inversion of the target F0 trajectory. This initialization provides an inversion-based starting point that stabilizes sampling under pitch-only guidance while keeping the reconstruction close to the model's data distribution. The proposed method requires neither melody-specific training nor modifications to the generative model. We demonstrate its applicability to both waveform- and latent-domain diffusion models for singing voice, guitar, and piano. Despite the nonlinear, non-Gaussian, and underdetermined nature of the task, our approach achieves melody adherence competitive with dedicated conditional baselines while retaining promising perceptual quality.
% }

\begin{figure*}[!t]
    \centering
    \includegraphics[width=\linewidth]{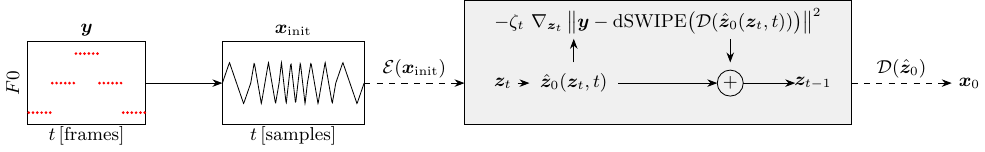}
    \caption{Overview of melody-guided sampling. The target F0 curve $\boldsymbol{y}$ synthesizes the warm-start signal $\boldsymbol{x}_\text{init}$, which is encoded to $\boldsymbol{z}_\text{init}$ and refined through guided reverse diffusion using the dSWIPE-based likelihood gradient.}
    \label{fig:melody_guidance_diagram}
\end{figure*}

We present a model-agnostic, training-free method for guiding unconditional diffusion models toward a target F0 trajectory, combining a differentiable pitch estimator as the measurement operator in DPS with the melody-informed warm initialization described above. An overview of our melody guidance method is shown in Figure~\ref{fig:melody_guidance_diagram}. The method requires neither melody-specific training nor modifications to the underlying generative model, and applies without modification to both waveform- and latent-domain diffusion models. We demonstrate it on singing voice, guitar, and piano. Despite the nonlinear, non-Gaussian, and underdetermined nature of the task, it achieves melody adherence competitive with dedicated conditional baselines while retaining promising perceptual quality.

This paper is structured as follows: Section \ref{sec:background} introduces DPS in waveform and latent domains. Section \ref{sec:method} presents the proposed melody-guidance method based on differentiable F0 estimation and warm initialization. Section \ref{sec:experiments} provides a hyperparameter analysis, audio examples, and comparisons with baseline methods. Finally Section \ref{sec:conclusion} provides some concluding remarks.

\label{sec:method}
\section{Background}
\label{sec:background}
\input{Background}

\section{Melody Guided Sampling}
\begin{figure}[!t]
    \centering
    \includegraphics[width=\linewidth]{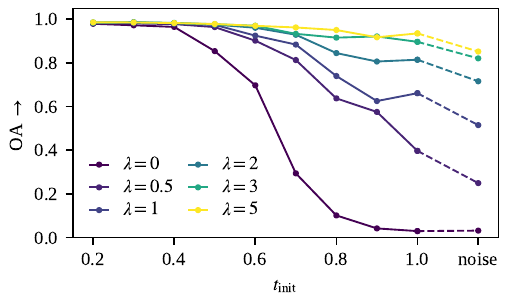}
    \caption{Mean OA as a function of $t_\text{init}$ across guidance strengths $\lambda$.}
    \label{fig:t_init_sweet_spot_oa_vs_triangle_all_lambda}
\end{figure}

\label{sec:method}
Figure~\ref{fig:melody_guidance_diagram} gives an overview of our melody guidance procedure. Given a target F0 trajectory $\boldsymbol{y}$, we synthesize a triangular wave $\boldsymbol{x}_{\mathrm{init}}$ to warm-start the reverse diffusion. Unlike a pure sine tone, it contains a harmonic structure that provides the diffusion model a more informative starting point, reducing the corrective guidance required during sampling. For latent diffusion models, this signal is encoded as $\boldsymbol{z}_{\mathrm{init}}=\mathcal{E}(\boldsymbol{x}_{\mathrm{init}})$. For waveform-domain models, we directly set $\boldsymbol{z}_{\mathrm{init}}=\boldsymbol{x}_{\mathrm{init}}$. We then start the reverse diffusion from it at time $t_\mathrm{init}$. At each denoising step, we guide the current sample toward the target melody by estimating the F0 of the decoded $\hat{\boldsymbol{z}}_0$ and comparing it to the target 
$\boldsymbol{y}$. This discrepancy provides a gradient that steers the next reverse-diffusion step toward audio following the prescribed melody.

We frame melody conditioning as an inverse problem, where we recover $\boldsymbol{x}_0$ given a target F0 trajectory treated as a degraded measurement $\boldsymbol{y} \approx \mathcal{A}(\boldsymbol{x}_0)$. We use the differentiable F0 estimator dSWIPE as the degradation operator $\mathcal{A}$. This formulation is challenging because an F0 curve is a nonlinear and highly underdetermined representation, as many waveforms with distinct timbres, articulations, and harmonic structures can share the same F0 trajectory. Thus, matching $\boldsymbol{y}$ alone does not uniquely determine a realistic audio signal. Following the framework described in \eqref{eq:dps_approx}, we approximate the likelihood-score gradient using dSWIPE yielding
\begin{equation}
     \nabla_{\boldsymbol{z}_t} \log p_t(\boldsymbol{y} \mid \boldsymbol{z}_t) \approx -\zeta_t\ \nabla_{\boldsymbol{z}_t} \left\| \boldsymbol{y} - \mathrm{dSWIPE}\big(\mathcal{D}(\hat{\boldsymbol{z}}_0(\boldsymbol{z}_t, t))\big) \right\|^2
    \label{eq:L_f0}
\end{equation}
with step size
$\zeta_t = {\lambda\, t \sqrt{N}} /{\left\|\nabla_{\boldsymbol{z}_t} \mathcal{L}\right\|^2}$, a noise-adaptive scaling shown to preserve manifold alignment during guided sampling~\cite{yang2024guidance}, where $ \mathcal{L}=\left\| \boldsymbol{y} - \mathrm{dSWIPE}\big(\mathcal{D}(\hat{\boldsymbol{z}}_0(\boldsymbol{z}_t, t))\big) \right\|^2$, $N$ the dimensionality of $\boldsymbol{z}_t$ and $\lambda$ a tunable guidance strength. To retain expressive melodic characteristics, such as vibrato, we apply Gaussian blurring to the target pitch representation $\boldsymbol{y}$ following~\cite{11044787}. Its standard deviation $\varsigma$, expressed in cents, controls the amount of pitch blurring.

We show in Section~\ref{sec:experiments} that relying on this guidance term alone yields an unnatural-sounding melody, and that high 
$\lambda$ values push the output out of the reference audio distribution. Warm starting with $x_\mathrm{init}$ reduces the amount of corrective guidance required during reverse diffusion, allowing us to use smaller $\lambda$ values while maintaining F0 adherence.
Using $\boldsymbol{x}_\text{init}$ alone can be viewed as timbre transfer from a triangular wave to a target instrument, as described in~\cite{andrade2026audio}. Nevertheless, our experiments in Section~\ref{sec:experiments} show that combining guidance with warm initialization is necessary to obtain melodically accurate and natural-sounding results. 
\begin{figure}[!t]
    \centering
    \includegraphics[width=\linewidth]{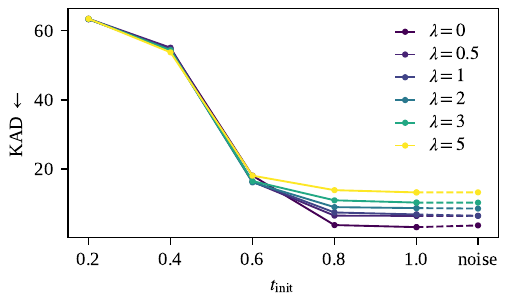}
    \caption{KAD as a function of $t_\text{init}$ across guidance strengths $\lambda$.}
    \label{fig:fad_tuning_grid_opensinger_noseed_n200_lines}
\end{figure}

\section{Experiments}
\label{sec:experiments}
In this section we evaluate our melody guidance method across four diffusion models, two waveform-domain piano models \cite{moliner2023solving, 11284630} and two latent diffusion models for singing voice and guitar \cite{svento2026loudar}. We restrict our baseline comparisons to publicly available methods with released implementations and pretrained checkpoints.
All audio examples can be found in the companion website \footnote{\href{https://cristobalandrade.github.io/MelodyGuidedGeneration/}{cristobalandrade.github.io/MelodyGuidedGeneration}}. 
 
\subsection{Effect of Guidance Strength $\lambda$ and Initialization Time $t_\text{init}$}
We begin by analyzing the effect of guidance strength $\lambda$ and initialization time $t_{\mathrm{init}}$ on melody adherence and audio quality. Here, $\lambda \geq 0$ scales the F0-matching term \eqref{eq:L_f0}, with $\lambda = 0$ disabling melody guidance and larger values pushing generation toward the target F0 trajectory at the risk of degraded audio quality \cite{moliner2023solving}. $t_{\mathrm{init}}$ sets how far into the reverse process generation begins. Larger values give the diffusion model more time to reshape the signal, at the cost of deviating from
$x_{\mathrm{init}}$'s melody, while smaller values preserve that melody more closely but leave the model less room to refine the tone. We compare two regimes. In the pure-noise regime, reverse diffusion starts from Gaussian noise. In the triangular warm-initialization regime, we synthesize $x_{\mathrm{init}}$ from the target F0 trajectory and initialize reverse diffusion at time $t_{\mathrm{init}}$. We evaluate the warm-initialization regime for $t_{\mathrm{init}} \in [0.2, 1]$ and sweep $\lambda \in [0, 5]$ jointly with $t_\text{init}$. For each configuration, we generate 100 samples using F0 trajectories from POP909 \cite{pop909-ismir2020} as targets and a LOUDAR checkpoint~\cite{svento2026loudar} trained on the GOAT guitar dataset~\cite{loth_goat_2025}, fixing $\varsigma = 50$ cents throughout. We report mean Overall Accuracy (OA)~\cite{raffel2014mireval}, computed using CREPE for F0 estimation, and Kernel Audio Distance (KAD)~\cite{kad} as proxies for melody adherence and audio quality, respectively.%in Figures~\ref{fig:t_init_sweet_spot_oa_vs_triangle_all_lambda} and~\ref{fig:fad_tuning_grid_opensinger_noseed_n200_lines}, respectively.

We first isolate the effect of $\lambda$ in the pure-noise regime. As shown in Fig.~\ref{fig:t_init_sweet_spot_oa_vs_triangle_all_lambda}, increasing $\lambda$ consistently improves OA. However, Fig.~\ref{fig:fad_tuning_grid_opensinger_noseed_n200_lines} shows that stronger guidance also increases KAD, indicating greater deviation from the reference audio distribution.

We then examine the effect of warm-starting the reverse diffusion with $x_\text{init}$ at decreasing $t_\text{init}$. Fig. \ref{fig:t_init_sweet_spot_oa_vs_triangle_all_lambda} shows that even with $\lambda = 0$, warm initialization alone substantially increases OA, and initializing later in the reverse process drives OA close to 1 regardless of $\lambda$. However, the same quality trade-off  reflected in the KAD persists.

Neither parameter alone reaches high melody adherence without degrading audio quality. This shows that guidance and warm initialization address complementary failure modes rather than substituting for one another, so both are needed jointly to achieve accurate and natural-sounding melody control. We therefore suggest a joint tuning strategy: start from a late warm init, increase $t_{\mathrm{init}}$ until melody hallucinations emerge, then raise $\lambda$ to suppress them.

\subsection{Baseline Comparison}
We build our melody-guided singing voice generation on the unconditional latent diffusion model LOUDAR, with a checkpoint trained on the OpenSinger dataset~\cite{hawthorne2022multi}, and guide its sampling process toward a target F0 trajectory as described in Section \ref{sec:method}, with $\lambda = 2.5$ and $t_\text{init} = 0.8$. For objective evaluation, we compare against DiffSVC~\cite{liu2021diffsvc}, a diffusion-based singing voice model conditioned on F0 and loudness curves, DiffSinger~\cite{liu2022diffsinger}, a diffusion model conditioned on a music score and WORLD~\cite{MoriseYO16}, a vocoder-based analysis-synthesis system with explicit F0 control, using random samples from OpenSinger as donors. As evaluation metrics, we report OA and Raw Pitch Accuracy (RPA) from \texttt{mir\_eval}~\cite{raffel2014mireval}, computed using both CREPE and SWIPE, alongside KAD \cite{kad} and Fr\'echet Audio Distance (FAD) \cite{kilgour2019frechetaudiodistancemetric} using the LAION-CLAP embedding provided in their respective tool kits \cite{kad, fadtk}.

We take the F0 from 100 melodies in POP909, and use these as guidance targets for all compared systems. Average metrics over this set are reported in Table~\ref{tab:all_baselines}. As the Singing Voice baselines were trained on different datasets, FAD and KAD values should be interpreted as indicative of relative audio quality rather than as directly comparable across systems. Informal listening suggests that our method achieves strong melody adherence while producing perceptually convincing singing voices. We note that the examples exhibit varying phonemes, which are resolved entirely by the unconditional model, since our method provides no control over them.

Next, we use LOUDAR trained on the guitar GOAT dataset~\cite{loth_goat_2025}, with $\lambda=2$ and $t_\text{init} = 0.7$, to compare our melody guidance against three baselines: RAVE~\cite{caillon2021ravevariationalautoencoderfast}, a VAE-based generator, trained on the ILLGuitarTimbre Dataset\footnote{\url{https://github.com/Intelligent-Instruments-Lab/IILGuitarTimbre}}, DDSP-Guitar~\cite{jonason2024ddspguitar}, a DDSP-based guitar synthesis model and Music Spectrogram Diffusion (MSD)~\cite{hawthorne2022multi}, a MIDI-conditioned diffusion model. Results are shown in Table~\ref{tab:all_baselines}. As in the singing voice setting, KAD and FAD values should be interpreted with caution, as all baselines were trained on different datasets. Informal listening suggests that our method achieves good F0 adherence and produces perceptually plausible guitar audio. However, some note onsets are insufficiently articulated, and string decays could be more pronounced. Onsets sometimes sound picked or bent, an articulation choice resolved by the unconditional model.

Finally, we apply melody guidance to CQT-Diff~\cite{moliner2023solving} and MRCQT-Diff~\cite{11284630}, two diffusion models operating on the CQT representation, both trained on the MAESTRO dataset~\cite{hawthorne2018enabling}, using $\lambda = 0.3$ and $t_\text{init} = 0.65$ for both models. We compare our approach against MSD and PianoKontext~\cite{gavrilev2026pianokontext}, a latent flow-matching model that renders variable-length piano performance segments given a deadpan latent context. As before, we report OA, RPA, KAD, and FAD.

Results are shown in Table~\ref{tab:all_baselines}. As all models in this setting were trained on the same dataset (MAESTRO), KAD and FAD values are directly comparable across systems, unlike in the singing voice setting. Our method achieves melody adherence competitive with dedicated conditional models, as reflected in the OA and RPA scores. 
Informal listening suggests that our method achieves F0 adherence but with degraded perceptual quality compared to the singing voice and guitar cases. Onsets are occasionally spurious or underarticulated, and decays lack piano-like characteristics. We attribute this to a distribution mismatch: the unconditional prior was trained on polyphonic MAESTRO recordings, making a monophonic melody line out of distribution and difficult for the diffusion prior to resolve.

\begin{table}[!t]
\centering
\fontsize{9}{10.8}\selectfont
\caption{Melody-guidance and baseline results per instrument. RPA/OA subscripts C and S denote CREPE and SWIPE pitch extraction, respectively.}
% \caption{Comparison across instruments and baselines, using 100 POP909 melodies per instrument. RPA/OA are reported as decimal fractions of melody MIDI ground truth, using CREPE (subscript C) and dSWIPE (subscript S) F0 front-ends. KAD/FAD use LAION-CLAP (music) embeddings, but their reference set differs by instrument and is \emph{not} comparable across blocks: for Singing Voice, each baseline is compared against its own training set; for Guitar, all methods are compared against a shared reference of real DI-guitar recordings from GOAT (2152 non-overlapping 10\,s clips); for Piano, all methods are compared against real MAESTRO clips, and are directly comparable since all piano models were trained on MAESTRO. For dSWIPE-based pitch metrics, note that dSWIPE cannot express ``unvoiced'' by construction, penalizing OA on ground-truth-silent frames, and that RPA$_{\text{S}}$ for MR-CQTdiff is partly self-referential, as dSWIPE also guided its generation. Best value per column in bold; second-best underlined where applicable.}
\label{tab:all_baselines}
\setlength{\tabcolsep}{4pt}
\begin{tabular}{lcccccc}
\toprule
Method & RPA$_{\text{C}}$ & OA$_{\text{C}}$ & RPA$_{\text{S}}$ & OA$_{\text{S}}$ & KAD\,$\downarrow$ & FAD\,$\downarrow$ \\
\midrule
\multicolumn{7}{c}{\textit{Singing Voice}} \\
\midrule

WORLD      & 0.59 & 0.66 & 0.61 & 0.46 & 27.6 & 0.48 \\
Diff-SVC   & 0.81 & 0.70 & 0.81 & 0.61 & 49.6 & 0.72 \\
DiffSinger & 0.76 & 0.85 & 0.76 & 0.68 & 29.6 & 0.30 \\
Ours       & \textbf{0.91} & \textbf{0.86} & \textbf{0.95} & \textbf{0.73} & \textbf{10.0} & \textbf{0.25} \\
\midrule
\multicolumn{7}{c}{\textit{Guitar}} \\
\midrule
RAVE                  & 0.92 & 0.75 & 0.94 & 0.71 & 27.4 & 0.43 \\
DDSP-Guitar   & \textbf{0.95} & \textbf{0.84} & 0.93 & 0.71 & 29.4 & 0.44 \\
MSD & 0.85 & 0.68 & 0.87 & 0.66 & 51.4 & 0.81 \\
Ours                     & 0.92 & 0.79 & \textbf{0.95 }& \textbf{0.72} & \textbf{18.6} & \textbf{0.32} \\
\midrule
\multicolumn{7}{c}{\textit{Piano}} \\
\midrule
PianoKontext         & 0.76 & 0.63 & 0.72 & 0.55 & \textbf{22.2} & \textbf{0.25} \\
MSD & 0.81 & 0.67 & 0.81 & 0.61 & 61.9 & 0.73 \\
Ours (MR-CQT)    & \textbf{0.96} & {0.78} & \textbf{0.99} & \textbf{0.78} & {29.5} & {0.39} \\
Ours (CQT)  & {0.95} & \textbf{0.88} & {0.97} & {0.74} & 52.4 & 0.66 \\
\bottomrule
\end{tabular}
\end{table}

% \subsection{Melody Guided Inpainting}
% Beyond timbre transfer and MIDI-to-audio rendering, our guidance mechanism also extends naturally to audio inpainting. Prior work has addressed inpainting in both the symbolic and waveform domains~\cite{min2023polyffusion, Moliner2024Diffusion}. We adopt the masked reconstruction term of~\cite{Moliner2024Diffusion}, $\mathcal{L}_{\text{INP}}(\boldsymbol{x}_{\text{obs}}, \boldsymbol{z}_t) = \nabla_{\boldsymbol{z}_t} \| \boldsymbol{x}_{\text{obs}} - \mathbf{m} \odot \hat{\boldsymbol{z}}_\theta(\boldsymbol{z}_t, t) \|^2$, where $\boldsymbol{x}_{\text{obs}}$ is the observed audio and $\mathbf{m}$ masks the known region. We add this reconstruction loss to the F0 consistency loss in \eqref{eq:L_f0} and use the gradient of the combined objective to guide generation.

% By constraining the masked region with a target F0 trajectory, rather than relying on the diffusion prior alone, our method fills gaps of up to 700 ms. This substantially exceeds the 200--300 ms gaps addressed in~\cite{Moliner2024Diffusion}. Informal listening indicates that combining both guidance terms yields more plausible completions than inpainting alone, though we leave a rigorous perceptual evaluation of this application to future work.

\section{Conclusion}
\label{sec:conclusion}
We presented a training-free method for guiding unconditionally trained diffusion models toward a target melody at inference time, formulating melody conditioning as an inverse problem solved via diffusion posterior sampling with a differentiable pitch estimator and a melody-informed warm initialization. Our experiments show that this simple, training-free approach is model-agnostic, applying without modification to both waveform- and latent-domain diffusion models, and achieves melody adherence competitive with dedicated conditional baselines across four models spanning piano, guitar, and singing voice, despite requiring no melody-specific training. Across these models, we observed different uncontrolled aspects of the generated audio, such as phoneme content, articulation, and note decay, emerging as artifacts of the unconditional prior, motivating future work on controlling them. Future work includes replacing the naive latent posterior sampling scheme with a learned latent degradation model to improve stability, extending the melody-informed warm initialization to other symbolic constraints beyond pitch, and applying the same guidance mechanism to long-gap audio inpainting constrained by a target pitch trajectory. A preliminary motivating result for this last direction is provided on the companion website.

\section{Acknowledgments}
We thank Eloi Moliner for providing MR-CQTDiff's checkpoint and for helpful discussions throughout this work. CA and SJS were supported through the joint German Academic Exchange Service (DAAD) and Research Council of Finland (RCF) Project (57763119) "Immersive Augmented Acoustics (IAA)". Part of the work was funded by the Deutsche Forschungsgemeinschaft (DFG, German Research Foundation) under Grant No. 500643750 (MU 2686/15-1). The International Audio Laboratories Erlangen are a joint institution of the Friedrich-Alexander-Universität Erlangen-Nürnberg (FAU) and Fraunhofer Institute for Integrated Circuits IIS.
{
\bibliographystyle{IEEEbib}
\bibliography{strings, refsFormatted}
}

\end{document}

%% file: Background.tex
\subsection{Diffusion Models}
Diffusion models learn to generate data by learning to transform a simple-to-sample distribution, typically Gaussian noise $\boldsymbol{x}_{1} \sim \mathcal{N}(\boldsymbol{0}, \boldsymbol{I})$ to the unknown distribution of the data $\boldsymbol{x}_0 \sim p_{0}$.
%distribution by reversing a gradual noising process. In the variance-exploding formulation, a clean sample $\boldsymbol{x}_0 \sim p_0$ is perturbed according to
% \begin{equation}
%     \boldsymbol{x}_t = \boldsymbol{x}_0 + \sigma(t)\boldsymbol{\epsilon},
%     \qquad
%     \boldsymbol{\epsilon}\sim\mathcal{N}(\boldsymbol{0},\boldsymbol{I}),
%     \label{eq:forward_diffusion}
% \end{equation}
The standard training procedure involves learning the score function $ \boldsymbol{s}_\theta(\boldsymbol{x}_t,t) \approx\nabla_{\boldsymbol{x}_{t}} \log p(\boldsymbol{x}_{t})$, at all levels $t$ of a stochastic process described by the
Itô SDE $ d\boldsymbol{x} = \boldsymbol{f}(\boldsymbol{x}_{t}, t)dt + g(t)d\mathbf{w}$

Following the parametrization proposed by \cite{karras2022elucidatingdesignspacediffusionbased} and considering the noise level $\sigma_t = t$, sampling from $p_0$ can be done via the Reverse ODE:
\begin{equation}
    d\boldsymbol{x}_t =
    -t\,\nabla_{\boldsymbol{x}_t}\log p_t(\boldsymbol{x}_t)\,dt.
    \label{eq:reverse_ode}
\end{equation}

At each diffusion step, the score also provides an estimate of the corresponding clean sample $\hat{\boldsymbol{x}}_0$ via the empirical posterior mean, $\hat{\boldsymbol{x}}_0 = \mathbb{E}[\boldsymbol{x}_0 | \boldsymbol{x}_{t}] \approx \boldsymbol{x}_{t} + t^2 \nabla_{\boldsymbol{x}_{t}} \log p_{t}(\boldsymbol{x}_{t}) $ which is a classic result of Tweedie's Formula \cite{efron2011tweedies}.
We use the denoised estimate $\hat{\boldsymbol{x}}_0$ to evaluate guidance objectives during sampling.

\subsection{Diffusion Posterior Sampling}
Many signal restoration tasks can be formulated as inverse problems. Given an unknown clean signal $\boldsymbol{x}_0$, an observation is modeled as
\begin{equation}
\boldsymbol{y}
=
\mathcal{A}(\boldsymbol{x}_0)
+
\boldsymbol{n},
\label{eq:inv_problem_formulation}
\end{equation}
where $\mathcal{A}$ is a known, possibly non-invertible degradation operator and $\boldsymbol{n}$ denotes observation noise. The objective is to recover samples that are both plausible under the data distribution and consistent with the degraded observation $\boldsymbol{y}$.

Diffusion posterior sampling addresses this task by sampling from the posterior distribution $p_t(\boldsymbol{x}_t\mid\boldsymbol{y})$. Using Bayes' Theorem its score decomposes as:
\begin{equation}
\nabla_{\boldsymbol{x}_t}
\log p_t(\boldsymbol{x}_t\mid\boldsymbol{y})
=
\nabla_{\boldsymbol{x}_t}\log p_t(\boldsymbol{x}_t)
+
\nabla_{\boldsymbol{x}_t}
\log p_t(\boldsymbol{y}\mid\boldsymbol{x}_t)
\label{eq:dps}
\end{equation}
Where the prior score is estimated by the pretrained diffusion model $\nabla_{\boldsymbol{x}_t}\log p_t(\boldsymbol{x}_t)$ and the likelihood score $\nabla_{\boldsymbol{x}_t}
\log p_t(\boldsymbol{y}\mid\boldsymbol{x}_t)$ enforces consistency with the observation.

The likelihood $
p_t(\boldsymbol{y}\mid\boldsymbol{x}_t)
=
\int
p(\boldsymbol{y}\mid\boldsymbol{x}_0)
p(\boldsymbol{x}_0\mid\boldsymbol{x}_t)
\, d\boldsymbol{x}_0$
is intractable because it requires marginalizing over the generally intractable conditional distribution $p(\boldsymbol{x}_0\mid\boldsymbol{x}_t)$. Diffusion posterior sampling \cite{chung2023diffusion} proposes to  approximate this via a surrogate function $p\left(
\boldsymbol{y}
\mid
\hat{\boldsymbol{x}}_0(\boldsymbol{x}_t)
\right)$ such that:%because $\boldsymbol{x}_t$ is a noisy state rather than a clean signal. In particular,
% \begin{equation}
% p_t(\boldsymbol{y}\mid\boldsymbol{x}_t)
% =
% \int
% p(\boldsymbol{y}\mid\boldsymbol{x}_0)
% p(\boldsymbol{x}_0\mid\boldsymbol{x}_t)
% d\boldsymbol{x}0.
% \label{eq}
% \end{equation}
\begin{equation}
\nabla_{\boldsymbol{x}_t}
\log p_t(\boldsymbol{y}\mid\boldsymbol{x}_t)
\approx
\nabla_{\boldsymbol{x}_t}
\log p\left(
\boldsymbol{y}
\mid
\hat{\boldsymbol{x}}_0(\boldsymbol{x}_t)
\right)
\label{eq}
\end{equation}

Additionally, considering the inverse problem formulation in \eqref{eq:inv_problem_formulation} the surrogate likelihood can be approximated by
\begin{equation}
\nabla_{\boldsymbol{x}_t}
\log p\left(
\boldsymbol{y}
\mid
\hat{\boldsymbol{x}}_0(\boldsymbol{x}_t)
\right)\approx  -\zeta_t
\nabla_{\boldsymbol{x}_t}\|
\boldsymbol{y}
-
\mathcal{A}(\hat{\boldsymbol{x}}_0)
\|_2^2.
\label{eq:dps_approx}
\end{equation}
where $\zeta_t$ is a noise-dependent scaling factor. This formulation accommodates both linear and nonlinear differentiable operators $\mathcal{A}$.

In our setting, $\boldsymbol{y}$ is a target F0 trajectory rather than a degraded waveform. Section \ref{sec:method} therefore introduces an F0-based likelihood-score that guides generation toward audio, whose estimated pitch contour matches a symbolic or user-specified melody.

\subsection{Latent Posterior Sampling}

Although waveforms are represented in a high-dimensional space, natural audio signals often lie near a substantially lower-dimensional manifold. Latent diffusion models exploit this structure by learning a compact representation with an autoencoder and performing the diffusion process in the resulting latent space. This reduces the computational cost of generation while retaining perceptually salient information~\cite{liu2023audioldm,evans2026stableaudio3}.

An encoder $\mathcal{E}$ maps a waveform $\boldsymbol{x}$ to a latent representation $\boldsymbol{z}=\mathcal{E}(\boldsymbol{x})$, and a decoder $\mathcal{D}$ reconstructs a waveform from the latent representation. A latent diffusion model estimates the score of the noisy latent distribution,
\begin{equation}
\boldsymbol{s}_\theta(\boldsymbol{z}_t,t)
\approx
\nabla{\boldsymbol{z}_t}\log p_t(\boldsymbol{z}_t)
\end{equation}
Analogously to the waveform-domain case, this score yields a denoised latent estimate, $\hat{\boldsymbol{z}}_0=
\boldsymbol{z}_t
+
t^2\boldsymbol{s}_\theta(\boldsymbol{z}_t,t)
$
whose corresponding waveform estimate is $\hat{\boldsymbol{x}}_0 = \mathcal{D}(\hat{\boldsymbol{z}}_0)$

Posterior sampling in the latent space follows the same principle as in waveform space. Given a condition $\boldsymbol{y}$, the likelihood gradient is evaluated through the decoded denoised estimate:
\begin{equation}
    \nabla_{\boldsymbol{z}_t}
    \log p_t(\boldsymbol{y}\mid\boldsymbol{z}_t)
    \approx
    \nabla_{\boldsymbol{z}_t}
    \log p\!\left(
        \boldsymbol{y}
        \mid
        \mathcal{D}(\hat{\boldsymbol{z}}_0)
    \right)
    \label{eq:latent_dps_approximation}
\end{equation}
This approach, commonly termed \textit{naive} latent posterior sampling, retains the training-free nature of posterior sampling, but requires back-propagating the guidance gradient through the decoder, which can be unstable \cite{Raphaeli_2025_ICCV}. In contrast, approaches that learn a degradation model directly in latent space avoid this but require additional training, and are therefore outside the scope of this work.
\subsection{Differentiable F0 Estimation}
Fundamental frequency (F0) estimation methods include classical signal-processing approaches, such as YIN~\cite{10.1121/1.1458024} and SWIPE~\cite{10.1121/1.2951592}, as well as neural approaches such as CREPE~\cite{kim2018crepe} and PESTO~\cite{pesto2023}. Using an F0 estimator in \eqref{eq:dps_approx} requires gradients that propagate from an F0-based condition to $\boldsymbol{x}_t$ at every diffusion step. We therefore use dSWIPE~\cite{11044787}, a differentiable formulation of SWIPE.

SWIPE estimates F0 per frame by comparing the input spectrum with sawtooth-inspired templates, each tied to a candidate frequency, then selects the highest-correlation frequency via a hard argmax. dSWIPE instead applies a softmax to the correlation scores, forming a distribution over pitch bins and yielding a differentiable expected pitch. In Section \ref{sec:method}, we use this output to define a melody likelihood-score that enables guiding waveform and latent diffusion models.
% For frame $m$, let $S_{\mathrm{sum}}(k,m)$ denote the score of candidate pitch bin $k$. The predicted pitch distribution is
% \begin{equation}
%     q_m(k)
%     =
%     \frac{
%         \exp\left(S_{\mathrm{sum}}(k,m)\right)
%     }{
%         \sum_{k'=1}^{K}
%         \exp\left(S_{\mathrm{sum}}(k',m)\right)
%     }
%     \label{eq:dswipe_distribution}
% \end{equation}
% Rather than producing a hard F0 estimate, Add Blurring!